# Bali Temple VR: The Virtual Reality based Application for the Digitalization of Balinese Temples

I Gede Mahendra Darmawiguna, Gede Aditra Pradnyana, I Gede Partha Sindu, I Putu Prayoga Susila Karimawan, Ni Kadek Risa Ariani Dwiasri

Laboratory of Cultural Informatics, Informatics Education Department, Universitas Pendidikan Ganesha
Singaraja, Bali, Indonesia
{mahendra.darmawiguna, gede.aditra, partha.sindu}@undiksha.ac.id, {yogasusila1996, risaariani01}@gmail.com

**Abstract**

The aim of this project is the development of a Virtual Reality Application in order to document one kind of Balinese cultural heritage which are Temples. The Bali Temple VR application will allow users to do a virtual tour and experience the landscape of the temples and all objects inside the temples. The application provides an on-site tour guide using virtual reality that allows users to experience the visualization of the Balinese cultural heritage, which in this case are temples. The users can walk through the temples and can see the 3D objects of temples and also there is narration of every object inside the temples with background music. Right now, the project has completed two temples for the virtual reality tour guide application. Those temples are Melanting Temples and Pulaki Temples. Based on the test results of its functional requirements, this virtual reality application has been able to run well as expected. All features that have been developed have been running well. Based on 20 respondents with various ages and backgrounds, our finding shows that The Bali Temple VR Application attracts people of all ages to use and experience it. They are eager to use it and hope that there will be more temples that they can experience to visit in this application.

# 1   Introduction

Bali is the smallest island and province of Indonesia. Bali, with its varied landscape of hills and mountains, rugged coastlines and sandy beaches, lush rice terraces and barren volcanic hillsides, all providing a picturesque backdrop to its colourful, deeply spiritual and unique culture, stakes a serious claim to be paradise on earth. Bali, the famed Island of the Gods has many cultural heritages, one of them are temples. Bali is known as an island of thousands temples. In every village in Bali, there are several temples and at least one small temple in each home of Balinese which reach to a total of 10.000. Balinese name of temple is pura. A pura is a Balinese Hindu temple and the place of worship for the adherents of Balinese Hinduism in Indonesia. Puras are built in accordance to rules, style, guidance and rituals found in Balinese architecture. Most of the puras are found on the island of Bali, as Hinduism is the predominant religion on it, however many puras exist in other parts of Indonesia where there are significant numbers of Balinese people. Puras are part of cultural heritages in Bali.

Cultural heritage is a cultural property that has an important value for understanding and developing history, science and culture in the framework of fostering the community and nation's personality [ThSR13]. Preservation of cultural heritage must continue to be carried out to preserve the culture that we have. Cultural heritage and natural resources are increasingly threatened by damage, not only by traditional causes, but because social change and socio-economic conditions are worsening the situation and as a great phenomenon of damage or destruction and awareness generally always comes too late for a cultural heritage, it's very important we take care.

Being located on the Pacific Ring of Fire (an area with a lot of tectonic activity), Indonesia has to cope with the constant risk of volcanic eruptions, earthquakes, floods and tsunamis. On several occasions during the last 15 years, Indonesia has made global headlines due to devastating natural disasters that resulted in the deaths of hundreds of thousands of human and animal lives, plus having a destructive effect on the land area (including infrastructure, and thus resulting in economic costs). The fact that Bali is part of Indonesia which is situated in the Pacific Ring of Fire, the efforts to prevent damage to cultural heritages need to be proactively managed by means of preservation of cultural heritages. One way to preserve the cultural heritage which in this case are temples is to use virtual reality technology. One of the main advantages, in the fields of virtual realities is the complete virtual reconstruction of 3D environments, especially buildings in full detail. Even if the cultural heritage is physically damaged, it can be virtually reconstructed through probabilistic approximations.

This paper presents the project for the development of a virtual reality based application for the documenting the Balinese temples in Bali, Indonesia. By having such a kind of model, we can reconstruct the buildings based on the data that have been developed in the virtual reality application if, for example, things unexpected happen in the future. The Bali Temple VR application will allow users to do the virtual tour and experience the landscape of the temples and all objects inside the temples.

**State of Arts: Laboratory of Cultural Informatics Augmented and Virtual Reality Projects.**

This laboratory focuses on research related to culture. One focus of research at this laboratory is the use of augmented technology and virtual reality to document the culture, cultural heritage, arts, folklore in Bali. Augmented and virtual reality technology development began in 2013. We first developed an augmented reality book containing Balinese cultural heritage such as temples, dances, musical instruments, building ornaments, and others [DKCW14][KDCA14]. The working concept of augmented reality is that the application will detect images in books that will then bring up immovable objects. Then in 2015, we developed an augmented reality book containing legend and Balinese folklore. The object in the application is an animation of every Balinese folklore scene [DSKA15]. The augmented reality project continues with applications without books. The feature used in the application to get a marker is to use a user-defined target, where the user uses images that are nearby which are then made into markers so that they can display animations from each scene [WiDS17]. The development of culture-based virtual reality at LCI began in 2018. The first virtual reality application developed were games for Balinese folklores [KhAP18]. Then, it was continued with the development of virtual reality applications for temples in Bali that had previously been developed using augmented reality on the augmented reality book project. The latest development of virtual reality application for Balinese temples is the development of a 3D 360° Virtual Reality Video Pura Besakih prototype [CrPW17].

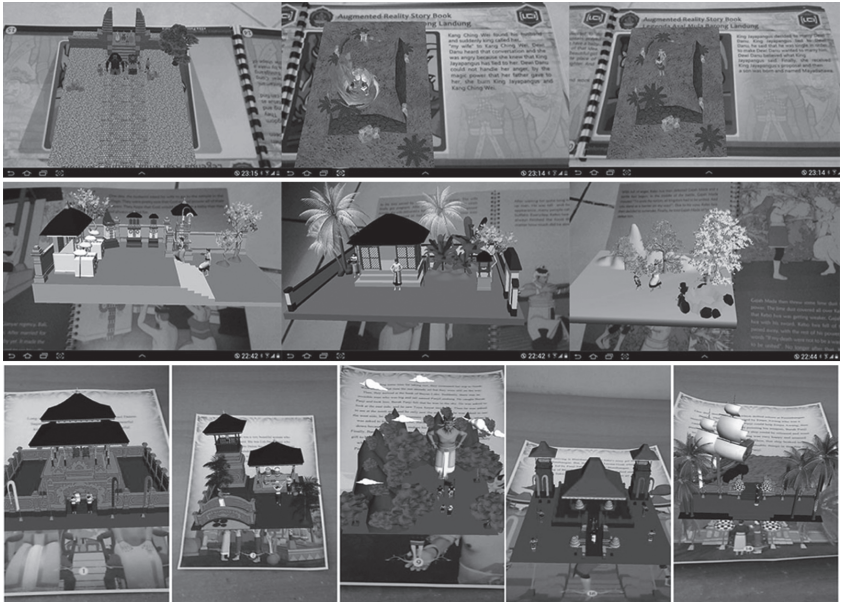

Fig. 1: Augmented Reality projects in Laboratory of Cultural Informatics

## 2   Materials and Methods

### 2.1   Analysis and Design
**Functional Requirements Analysis**
After gathering and analyzing needs, there are some functional requirements that are used as a basis for designing Bali Temple VR applications, which are:
  a. The application facilitates users to be able to take a tour in the temple area.
  b. Applications can display temple objects and their landscapes in three dimensions.

    c. The application can provide information in the form of audio and text about objects in the temples.

**Tools and Materials**

Based on the needs analysis carried out, the software that is needed for the development of virtual reality applications is:
- Blender is used to create Pulaki Temple objects
- Photoshop is used to edit temple object textures
- Unity3D with StemVR Library is used to create the virtual reality application
- Adobe Premiere is used to edit audio for temple object explanations.

The hardware that was used for the development and implementation of this application is:
- PCs for developing and implementing the application.
- HTC Vive

## 2.2 The Development Steps of Bali Temple VR Application

The steps in developing the Bali Temple VR consist of three main steps. First, taking real images of temples including the objects inside the temples by using a drone and perspective calculation in order to get the exact measurement. Second step is the development of 3D objects of the temples by using Blender. The third step is the development of VR application by using Unity3D and the library Stem VR. The VR tool that is used is HTC Vive.

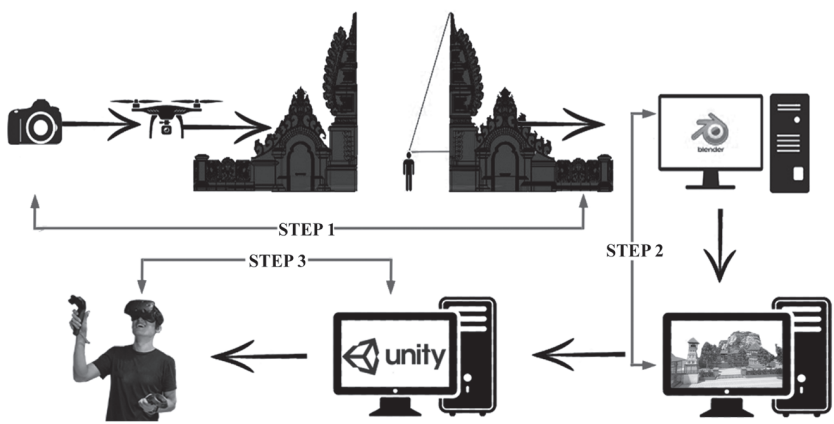

Fig. 2: The steps of Virtual Reality Application Development

# 3 Results

## 3.1 Implementation of 3-Dimensional Objects

The first stage in the development of the augmented reality story book is the creation of 3D objects and animations. The tool used is Blender version 2.70. The development of 3D objects and animation involves several steps:

a. **Modeling.** This is the process of creating 3D mesh (characters and other 3D objects) using some mesh tools such as plane, cube, circle, uv sphere, icosphere, cylinder, cond, grid, and torus. The techniques that were used were extrude technique and mirror modification.

b. **Materials & Texturing.** This process is used to simulate a surface color or property of 3D mesh. This is based on a photorealistic interpretation of a real material or any materials that can produce similar surface colors and textures.

c. After the objects and the landscapes are done they are exported to Unity3D.

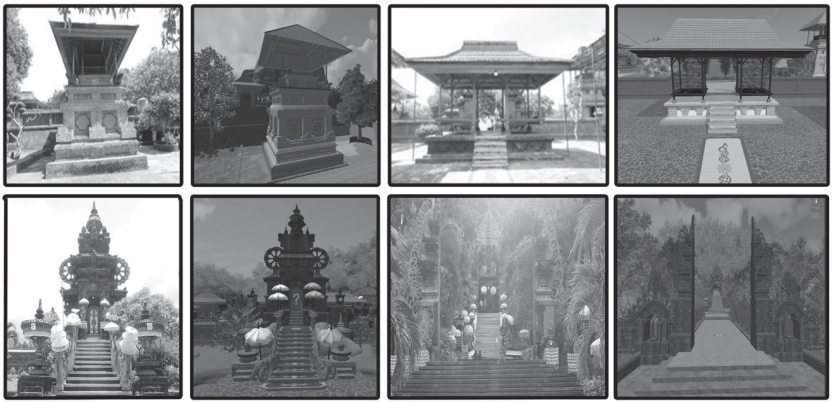

Fig. 3: Comparison between real Puras' objects and 3D Objects developed in Blender

## 3.2 Implementation of Virtual Reality Application

The development of the virtual reality application uses Unity3D. Before working with Unity3D, Steam VR has to be installed in order to use the virtual reality features. Steam VR can be downloaded on the Unity website, https://assetstore.unity.com/.

The steps for creating a virtual reality application are as follows:
  a. Several files must be imported to Unity3D. The files are 3D objects with format (.fbx) and the texture files of the entire objects.
  b. It is necessary to create a number of scenes in Unity3D, and then those scenes are stored in the Asset folder. The number of scenes depends on the number of temples. For example, the temple area is divided into five parts, meaning that five scenes are made. Every scene on Unity3D consists of Camera Righ, Directional Light, Steam VR Laser Pointer and Steam VR Teleporter.
  c. To run the teleporter in Unity, a mesh collider is set on the object to be targeted by the teleporter. For example, in the VR application that was developed, the temple page was used as the teleport target so users could walk around the temple area by directing the controller to the temple page.
  d. The final step is the process of building a Virtual Reality application. Application settings can be set in the build file menu settings. In this setting you can select the scene to be built and you can also add application icons. Then, the Virtual Reality application is ready to be built and run. The application results will be in the apk or exe format.

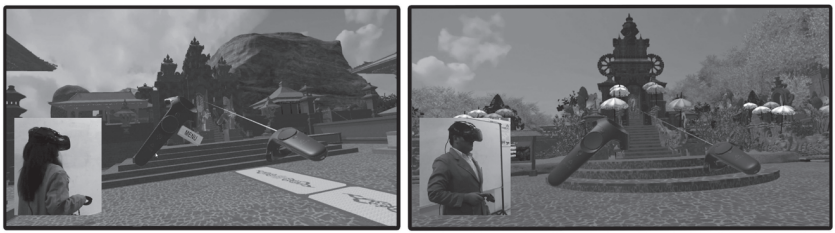

Fig. 4: The Bali Temple VR Application

## 4	Discussion

The application gives an on-site tour using virtual reality that allows users experience the visualization of the Balinese cultural heritage, which in this case are temples. The users can walk through the temples and can see the 3D objects of temples and also there is narration of every object inside the temples with background music.

There are several tests we have done. Based on the test results of its functional requirements (black box testing), this virtual reality application has been able to run well as expected. All features that have been developed have been running well. We also tested the application with a content expert to test the accuracy of the shape of each object temple created in the virtual reality application. The accuracy of the objects and landscapes are 90%. There are several objects that need to be polished.

In order to get the response from the people, the application is placed in the Museum Bali. The questionnaire that is used is a User Experience Questionnaire (UEQ). UEQ is used to measure the user experience of interactive products. The scale of the questionnaire cover both classical usability aspects (efficiency, perspicuity, dependability) and user experience aspects (originality, stimulation).

| Item | Mean | Variance | Std. Dev. | No. | Left | Right | Scale |
|---|---|---|---|---|---|---|---|
| 1 | 2.1 | 0.1 | 0.3 | 20 | annoying | enjoyable | Attractiveness |
| 2 | 2.1 | 0.1 | 0.2 | 20 | not understandable | understandable | Perspicuity |
| 3 | 2.1 | 0.1 | 0.3 | 20 | creative | dull | Novelty |
| 4 | 2.1 | 0.1 | 0.3 | 20 | easy to learn | difficult to learn | Perspicuity |
| 5 | 2.3 | 0.2 | 0.5 | 20 | valuable | inferior | Stimulation |
| 6 | 2.1 | 0.1 | 0.3 | 20 | boring | exciting | Stimulation |
| 7 | 2.5 | 0.3 | 0.5 | 20 | not interesting | interesting | Stimulation |
| 8 | 1.4 | 0.6 | 0.8 | 20 | unpredictable | predictable | Dependability |
| 9 | 1.9 | 0.1 | 0.3 | 20 | fast | slow | Efficiency |
| 10 | 2.1 | 0.1 | 0.2 | 20 | inventive | conventional | Novelty |

| Item | Mean | Variance | Std. Dev. | No. | Left | Right | Scale |
|---|---|---|---|---|---|---|---|
| 11 | 2.1 | 0.1 | 0.3 | 20 | obstructive | supportive | Dependability |
| 12 | 2.1 | 0.7 | 0.8 | 20 | good | bad | Attractiveness |
| 13 | -0.1 | 0.7 | 0.9 | 20 | complicated | easy | Perspicuity |
| 14 | 2.0 | 0.0 | 0.0 | 20 | unlikable | pleasing | Attractiveness |
| 15 | 2.4 | 0.2 | 0.5 | 20 | usual | leading edge | Novelty |
| 16 | 2.1 | 0.1 | 0.2 | 20 | unpleasant | pleasant | Attractiveness |
| 17 | 2.1 | 0.1 | 0.3 | 20 | secure | not secure | Dependability |
| 18 | 2.1 | 0.1 | 0.3 | 20 | motivating | demotivating | Stimulation |
| 19 | 2.0 | 0.0 | 0.0 | 20 | meets expectations | does not meet expectations | Dependability |
| 20 | 2.1 | 0.1 | 0.3 | 20 | inefficient | efficient | Efficiency |
| 21 | 2.2 | 0.2 | 0.4 | 20 | clear | confusing | Perspicuity |
| 22 | 2.1 | 0.1 | 0.2 | 20 | impractical | practical | Efficiency |
| 23 | 2.1 | 0.1 | 0.2 | 20 | organized | cluttered | Efficiency |
| 24 | 2.1 | 0.1 | 0.2 | 20 | attractive | unattractive | Attractiveness |
| 25 | 2.2 | 0.1 | 0.4 | 20 | friendly | unfriendly | Attractiveness |
| 26 | 2.4 | 0.2 | 0.5 | 20 | conservative | innovative | Novelty |

Table 1: Result of UEQ for User Experience Test.

Based on 20 respondents with various ages and backgrounds, our finding shows that The Bali Temple VR Application has positive impressions in groups in Attractiveness, Clarity, Efficiency, Accuracy, Stimulation and Novelty. The Bali Temple VR Application attracts people of all ages to use and experience it. They gave a very positive response to the project that had been developed. They are eager to use it and hope that there will be more temples that they can experience to visit in this application.

## 5    Conclusion

The Bali Temple VR Project gives the opportunity to document cultural heritage in Bali where the main goal of this project is to preserve Balinese cultural heritage and to prevent loss of critical information when negative things like natural disasters occur in the future. In addition, this application can be used as a tool to introduce the culture of Bali to the people who are interested to know more. The Bali Temple VR application can help users to find information about Balinese temples as cultural heritage. This application can be used as a documentation for cultural heritage for the Bali Provincial Cultural Service. This application can also help to introduce Pura on Bali globally. Currently, only two temples have been digitized in the form of virtual reality applications and there are three temples that are under construction. This project is carried out with self funding from researchers. Hopefully, there will be a form of funding in the future that can help us develop virtual reality applications for digitally massive cultural heritage in Bali.

## Literature

[CrPW17]   Crisnapati, Padma Nyoman; Prihantana, Made Agus Suryadarma and Wijaya, Bagus Kusuma. "Pengembangan Prototipe 3D 360° Virtual Reality Video Pura Besakih Menggunakan Blender." E-Proceedings KNS&I STIKOM Bali (2017): pp. 437–440.

[DKCW14]   Darmawiguna, I.G.M.; Kesiman, M.W.A.; Crisnapati, P.N.,; Wiartika, I.M.E.; Suparianta, K.D.; Susena, I.K.; Yudiantara, I.M.: Augmented Reality for the Documentation of Cultural Heritage Building Modelling in Bali, Indonesia. Kultur un Informatik: Reality and Virtuality, Proc of Culture and Computer Science Conf, Berlin, Germany, 2014, pp. 107–117.

[DSKA15]   Darmawiguna, I. Gede Mahendra; Sunarya, I. Made Gede; Kesiman, Made Windu Antara; Arthana, Ketut Resika and Crisnapati, Padma Nyoman: The augmented reality story book project: A collection of balinese myths and legends. In International Conference on Augmented and Virtual Reality, pp. 71–88. Springer, Cham, 2015.

[KDCA14]　Kesiman, M.W.A., Darmawiguna, I.G.M., Crisnapati, P.Y., Ardipa, G.S., Mariyantoni, I.K.Y., Nugraha, M.L., Prasetya, A.Y.R.A., Dewantara, I.M.A.Y. : The AR Book Project: Collection of Augmented Reality Application of Balinese Artistic and Cultural Objects. Kultur un Informatik: Reality and Virtuality, Proc of Culture and Computer Science Conf, Berlin, Germany, 2014, pp. 93–105.

[KhAP18]　Khoerniawan, R.W., Agustini, K., Putrama., I. M : Game Edukasi Penjelajah berbasis Virtual Reality. Kumpulan Artikel Mahasiswa Pendidikan Teknik Informatika (KARMAPATI), Volume 7, No. 1, 2018.

[ThSR13]　Threesiana, R.; Suwardhi, D.; Riyanto, S: Development Of Virtual Reality GIS For Cultural Heritage Conservation (Case Study: Sewu Temple). Indonesian Journal of Geospatial, ISSN: 2089-5054, Volume 2, No. 1, 2013.

[WiDS17]　Wiradarma, I. Gusti Gede Raka; Darmawiguna, I. Gede Mahendra and Sunarya, I. Made Gede: Pengembangan Aplikasi Markerless Augmented Reality Balinese Story "I Gede Basur". Jurnal Nasional Pendidikan Teknik Informatika (JANAPATI) 6, no. 1 (2017): pp. 30–38.